\documentclass[structabstract,letterpaper]{aa}  

\usepackage{graphicx}
\usepackage{psfig}
\usepackage{epsfig}
\usepackage{enumerate}
\usepackage{multirow}

\usepackage{listings}
\usepackage{txfonts}
\usepackage{epstopdf}

\newcommand{\be}{\begin{equation}}
\newcommand{\ee}{\end{equation}}
\newcommand{\ba}{\begin{eqnarray}}
\newcommand{\ea}{\end{eqnarray}}

\def\gs{\mathrel{\raise1.16pt\hbox{$>$}\kern-7.0pt %
\lower3.06pt\hbox{{$\scriptstyle \sim$}}}}         %
\def\ls{\mathrel{\raise1.16pt\hbox{$<$}\kern-7.0pt %
\lower3.06pt\hbox{{$\scriptstyle \sim$}}}}         %

\begin{document}

\title{Cloud Cosmology : Building the Web Interface for iCosmo}

\author{T. D. Kitching\inst{1}, A. Amara\inst{2}, A. Rassat\inst{3}, A. Refregier\inst{3}}

\institute{
  Oxford Astrophysics, Department of Physics, Keble Road, Oxford, OX1 3RH, United Kingdom
  \and
  Department of Physics, ETH Zurich, Wolfgang-Pauli-Strasse 16, CH-8093 Zurich, Switzerland
  \and
  IRFU-SAP, Service d'Astrophysique, CEA-Saclay, F-91191 Gif sur Yvette Cedex, France
  } 

\date{Submitted: \today}

  \abstract
   {}
   {Astronomy and cosmology have embraced the internet. We routinely 
     and successfully use the 
     internet as a repository for sharing code, publications  
     and information, and as a computational resource. 
     However the interactive nature of the 
     web, for use as an alternative to downloading code has been largely overlooked. 
     In this article we will outline a simple framework in which a cosmological 
     code can be turned into an interactive web interface. This is presented as a result 
     of creating {{\tt www.icosmo.org}} which is a front-end for the open-source 
     software {\tt iCosmo}.}
   {We explain how an HTML page can be created and how 
     a cosmological code can be incorporated into a web environment using CGI scripts. We 
     outline how plots and downloadable text can be made, and describe how help and 
     documentation can be created.}
   {By using simple HTML and CGI scripts a basic web interface for any 
     cosmological code can be created easily. We provide a worked example of the 
     methods outlined, which can be used as a simple template by any researcher who wants 
     to share their work online.}
   {}

   \keywords{cosmology --
             observables --
             numerical methods}

   \maketitle

\section{Introduction}
\label{Introduction}
The internet has begun to revolutionise the way that astronomical research is conducted. 
Online paper archives (e.g the arXiv\footnote{{\tt http://arxiv.org/}}) have allowed 
for 
the dissemination of articles and results, and data 
(e.g. SDSS\footnote{{\tt http://www.sdss.org/}} Adelman-McCarthy et al., 2006) 
are routinely made publicly available 
online for anyone to analyse. We of course use the internet to communicate via email, voice 
and video connections. More recently 
there have been some efforts to use the resources of the internet to aid in the 
calculation of a specific suite of scientific results; for example the Galaxy Zoo\footnote{{\tt 
http://www.galaxyzoo.org/}} 
(Lintott et al. 2008) and cosmology@home\footnote{{\tt 
http://www.cosmologyathome.org/}} both of which solve particular problems 
of concern for the authors and simultaneously serve as remarkable outreach programmes. 

The open source code made available through the internet has 
enabled cosmologists to use this 
code, saving their time to do further research. Of particular note are the CMB 
power spectrum codes {\tt CMBfast} (Zaldarriaga \& Seljak, 2000) and 
{\tt CAMB} (Lewis et al., 2000) and some likelihood analysis codes for example 
COSMOMC (Lewis \& Bridle, 2002) and BAYESYS\footnote{{\tt http://www.inference.phy.cam.ac.uk/bayesys/}}. 

However there have been relatively few online resources that cosmologists can use 
in their research as a replacement to downloading code. 
Notable and excellent exceptions include the CMB 
LAMBDA\footnote{\tt http://lambda.gsfc.nasa.gov/toolbox} toolbox, and the impressive 
online data reduction
interface AstroWISE\footnote{\tt http://www.astro-wise.org/}, there are also some basic
cosmology calculators (e.g. Wright, 2006). 
The use of the internet in this fashion 
heralds the move away from using the internet as a computational farm (a source of 
CPU's or people used for a specific suite of problems), 
or as a simple repository for data or code, towards a mode in which the 
interactivity of the internet is harnessed to make the act of performing useful 
cosmological calculations an online, interactive experience. 
Cloud computing is the term that is given 
to the process of using a non-local anonymous machine to perform computational tasks. 
We call the online computation of cosmological products \emph{cloud cosmology}.  

Cosmological calculations were originally done using simply paper and pencil. The advent 
of the computer made more complex calculations quicker and easier to compute. 
The internet has enabled code to be shared between the community, and  
the next step in this evolution is so create an interface 
for the code itself on the internet. 

The danger in creating web-interfaces for complex code is that a `black box' 
mentality is encouraged. In this article we strongly advocate the opposite 
approach and work within the philosophy that everything created should be a 
`transparent box' by which we mean that 
the code powering the site should always be made available and that 
documentation explaining the calculations and the code should be provided. 

The aim of this article is to guide the reader 
through the process of turning an open source code into a web interface. 
This will be done within the context of presenting  
the interactive tools on {\tt http://www.icosmo.org} that we have created as a front-end 
web interface for the open source code {\tt iCosmo} (Refregier et al., 2008). 
The website allows cosmological distances, Hubble parameter, growth factor, 
linear and non-linear matter power spectra, 
cosmic shear power spectra, baryon distance scales, supernovae 
magnitude distances; and lensing, BAO and supernovae Fisher matrices to be calculated 
online in real-time.  
In addition to the interactive web pages, 
there are links to the 
open source {\tt iCosmo} code and to some tutorial/teaching resources 
for cosmology and the probes used to measure cosmology. 

In this article, we will not describe all possible ways that a web interface could 
be created. We will present the process from the perspective of cosmologists 
(who are not experts in web design) who want to make their code 
easily accessible to a wide audience and want an easy and quick way to make code 
available.
We will begin by describing the requirements that a working cosmologist may need from 
an interactive web interface. Then we will describe the particular solution that we have 
used, along with some alternatives that we have investigated along the way. This will be
written in a pedagogical style from the point of view of a non-expert 
in computer science. 

The {\tt iCosmo} HTML and CGI will be released in {\tt iCosmo} v1.2 
({\tt http://icosmo.pbwiki.com}). A description of the interactive features of the website are given in Appendix A.

\section{Web solutions}
\label{Web solutions}
The needs of the cosmological community are not entirely unique to science but are 
fairly representative in terms of the types of product that are required from a 
useful web interface. We will explain the requirements that we need as cosmologists and 
then step through the solutions that we have found. 

\subsection{Requirements from Cosmology}
\label{Requirements from Cosmology}
To be a true replacement to downloading code the web interface must be able to perform 
the majority of the actions, and create all the products, 
that one may wish to make using the code. 

A user must be able to change the inputs easily. In cosmology these are commonly values 
of cosmological parameters. There are also some numerical limits and resolutions of 
particular variables on which cosmological functions depend; for example redshift $z$, 
wavenumber $k$, azimuthal wavenumber $\ell$.

The two most useful end-point products are the plot (or graph) that can be used to 
visually represent some functional behaviour, and a table of the data in the graph 
in a format that can then be used in subsequent programmes. The graphical interface 
should be flexible enough that any aspect of the graph can be tailored online to the 
users needs. In cosmology we often need to plot 
functions for multiple values of cosmological parameters including error bars. 

The {\tt iCosmo} code (Refregier et al., 2008) is a low-redshift, 
dark Universe, cosmology tool that 
allows a user to create cosmological functions, observables and parameter forecasts for 
arbitrary cosmologies and experimental designs. Our task was to turn this into an 
interactive web interface. 

A simple demonstration of the procedures covered in this Section are shown on 
\begin{itemize}
\item {\tt http://www.icosmo.org/WebPaper.html}
\item {\tt http://www.icosmo.org/cgi-bin/WebPaper.cgi}
\end{itemize}
and in Appendix B, these examples can be used as a template to create new cosmological 
web interfaces.

\subsection{HTML}
\label{HTML}
HTML (HyperText Markup Language) is the language in which web pages are written. 
A website consists of a group of pages, usually in some directory structure. 
HTML is akin to the familiar {\tt Latex} 
typesetting language in which a script is written and then compiled to create a 
document. In this case however it is the web browser that does the compiling when a 
particular page is loaded. HTML pages can be created in most word-processing or text 
based programmes (e.g. MS Word, Emacs, etc.) and just save the file with the 
extension {\tt .html}. The webpage can then be viewed by typing in the path to 
the file. Note that when a browser calls a webpage it does so as an external user, so 
permissions on any files need to be such that a global user can execute the HTML 
code (e.g. in UNIX using the {\tt chmod} command). In general one must be careful about 
permissions, if a file is \emph{writable} by a global user then a malicious user could turn
such a file into a virus. However if a file is writable but not executable this makes 
such an attack much more difficult, in developing {\tt iCosmo} we followed the 
guidelines set out on this webpage 
{\tt http://www.oucs.ox.ac.uk/web/
faq/index.xml.ID=safeperl}.

A very simple HTML page is shown here : 
\\
\\
\noindent {\tt <HTML>\\
<head>\\
<title>Hello World</title>\\
</head>\\
<body>\\
<font color="yellow">\\
 Hello World\\
</font>\\
</body>\\
</HTML>\\}
\\
\noindent The first line declares that this is an HTML code, and the last line 
is the end the code. The 
{\tt head} contains information that will not appear on the page itself but may affect its 
attributes. The {\tt body} contains information that will appear on the page, here there
is a simple style change made to the font of the text shown. We have found that the 
following pages were particularly useful in documenting the HTML syntax
\begin{itemize}
\item {\tt http://www.w3schools.com/html/DEFAULT.asp}
\item {\tt http://www.htmlgoodies.com/}
\item {\tt http://www.htmlcodetutorial.com/}
\item {\tt http://en.wikipedia.org/wiki/HTML}
\end{itemize}
though this is by no means an exhaustive list and typing in HTML and a problem into any search engine usually results in finding the solution very quickly. To check that the HTML page that you have written is valid (no bugs) there is an online HTML compiler that will highlight any problems or inconsistencies at {\tt http://validator.w3.org/} -- to 
use this the website needs to be online.

Writing `bare' HTML allows the webpage to be exactly tailored to your needs, however it 
is a time consuming activity and the syntax needed to create complex displays can become
cumbersome. Thankfully there are some very good website creators available that allow 
websites to be created in much the same way that a ``Power Point'' (or Keynote) 
presentation is made e.g. creating texboxes, changing fonts and colours, adding 
pictures etc. For the front end of the {\tt iCosmo} site we used {\tt iWeb} for Mac OSX, 
for Windows there is {\tt FrontPage}, and there are even online webpage creators such as 
Google Sites. 

To enable more advanced features in a webpage one has to use scripts. There are 
different types of script, those that can be run within HTML itself and those 
that cannot. In {\tt iCosmo} we have used some in-HTML JAVA scripts to enable some 
advanced features, for example the tabbed environment (see Appendix A) and some more 
simple things like allowing all checkboxes to be selected at once (see Section 
\ref{Input and Output}). These in-HTML scripts are defined in the {\tt <head>} section of 
the HTML and are initialised using {\tt <script language=javascript>} and ended 
using {\tt </script>}, there are many open-source javascript widgets which can be copied 
and pasted from the web into any HTML page. 

\subsection{Running Scripts}
In creating a web interface for cosmology one will inevitably have to run some sort of 
programme that will take inputs (say cosmological parameters) and perform a cosmological 
calculation. Running an executable, or script, can be done either on the side of the server (where the webpage is stored, see Section \ref{Servers}) or can be done through the browser itself on the `client' side. We opted for, and recommend, the server-side approach 
since cosmological calculations can be complex (a clients computational power may be minimal) and server side scripts are safer for the client (they do not need to run a potentially dangerous code on their personal machine).

There are a number of ways which we encountered that allow 
scripts to be run. Most simply any C code can be compiled which can be run 
in a browser by typing the 
executable name into the command line, though creating an entire webpage out of a C code 
quickly becomes complex. Alternatively one can use JAVA\footnote{{\tt 
http://java.com/en/}} 
and then create a ``Applet'' (although this is a client side solution). Other 
alternatives include PHP and API which are additions to HTML that allow scripts 
to be run from within HTML. An example C code, compiled like {\tt gcc test.c -o test.cgi} is given below: 
\\
\\
\noindent {\tt 1 \#include <stdio.h>\\
2 int main(void) \{ \\
3 float p,q,r; \\
4 p=2.*2.;\\
5 q=2.*2.*.2;\\
6 r=p+q;\\
7 printf("Content-Type: text/plain; charset=us-ascii$\backslash$n$\backslash$n");\\
8 printf("Hello world$\backslash$n$\backslash$n");\\
9 printf("2*2+2*2*2=\%f$\backslash$n",r);\\
10 return 0;\\
11 \}\\
}
\\
\\
The easiest solution that we have found is to use CGI (Common Gateway Interface) scripts. 
These are executable scripts (akin to a more complicated version of a shell script) that 
can be written in a variety of different code languages. When a CGI script's URL is 
called the server runs the CGI script, that creates a web page which is then displayed. 
The most common language in which CGI scripts are written, and the one which we have 
chosen for {\tt iCosmo}, is PERL\footnote{{\tt http://www.perl.com}}. 
For {\tt iCosmo} we chose PERL over Shell scripts because PERL is a full 
programming language with many available libraries, and as such is somewhat 
more flexible.
An example PERL CGI script is shown here:
\\
\\
\noindent {\tt \#!/usr/bin/perl\\
print<<HTML;\\
<HTML>\\
<head>\\
<title>Hello World</title>\\
</head>\\
<body>\\
<font color="yellow">\\
 Hello World\\
</font>\\
HTML\\
\#this is a comment\\
\#do a simple calculation in PERL\\
\$x=2*2+2*2*2;\\
\#display the calculation on the webpage\\
print<<HTML;\\
2<sup>2</sup>+2<sup>3</sup>=\$x\\
HTML\\
\#end the HTML\\
print<<HTML;\\
</body>\\
</HTML>\\
HTML\\
exit;\\
}\\ 
\\
\\
The first line declares that this is a PERL script. The {\tt print<<HTML} then says 
``the following is HTML'' where the header is created. Calculations can then be done in 
PERL and displayed on the webpage, in this very simple example we calculated $2^2+2^3$. 
We also show here an example CGI shell script: 
\\
\\
\noindent {\tt 
1 \#!/bin/sh\\
2 \# xmltest.sh\\
3 \# XML-based "Hello World"\\
4 echo Content-type: text/xml\\
5 echo\\
6 echo "<?xml version=$\backslash$"1.0$\backslash$" encoding=$\backslash$"UTF-8$\backslash$"?>"\\
7 echo "<message>"\\
8 echo "<text>Hello World!</text>"\\
9 echo "</message>"\\
10 exit 0\\
}
\\
\\

\subsection{Servers, URLs, IP Addresses}
\label{Servers}
A webpage is hosted on a \emph{server} which, as far as we are concerned, 
is essentially a computer with some specific software installed that allows 
CGI scripts to be run, and on which the web site resides. 
When you click on the URL of any webpage the browser will be redirected 
to a server, for a simple webpage the browser will download the webpage HTML from the 
server and display it, for a CGI script the server will run the script and then 
send the HTML created to the users browser. 
Servers are nothing more than a normal computers 
with extra software installed, indeed even a laptop can become a server 
(this is actually very easy with some recent operating systems e.g. Mac OSX Leopard, 
where server software is included as standard). The most common form of server software is 
Apache\footnote{{\tt http://www.apache.org/}}. 
A webpage on a server typically consists of a number of 
folders : one containing the static pages (front page of website etc.), 
one containing the CGI scripts, one containing some shared files. 

For {\tt iCosmo} we developed the site using Apache installed on a MacBook laptop, 
to turn the laptop into a server all one needs to do is turn on ``web sharing'' 
once Apache is installed. Later we transfered the site to a dedicated server. 

Every computer and every server connected to the internet has a unique numerical 
indentifier called an IP (Internet Protocol) address. To connect to a 
servers website from a browser the IP address of the server needs to be known. 
The key thing to realise here is that IP addresses (computers/servers) use ports, 
through which they communicate.
When two computers interact the IP address says where the machine is, 
the port says what type of interaction will occur. HTML web pages use port 80. 
So the server needs to have port 80 open to allow IP addresses to accesses HTML pages 
stored on the server.

If a computer is connected to a network then it may have an internal IP for the 
network 
and the router will have an external IP address. 
When developing a website at home, 
through a router for instance, the routers software should be changed to open port 80. 
There is a very useful website here that will provide a step-by-step guide to do this for 
almost any router available {\tt http://www.PortForward.com}, the router then needs to 
re-route any incoming traffic through the port to the internal IP of the machine. 
Once port 80 is open then typing in the external IP to a web browser will take 
the browser to your servers webpage.

A URL (Uniform Resource Locator) is an alpha-numeric replacement for an IP 
address. When developing a website you will be able to use the IP as an address for the 
webpage e.g. 
typing {\tt http://193.168.0.1} will take you to your webpage. If you want to create 
a URL instead of using the IP address then DNS (Domain Name System) 
server can be used which will reroute any traffic using a given URL to your IP address. 
There are 
a number of free services that do this, when developing {\tt iCosmo} we used DynDNS 
{\tt www.dyndns.com} which is a dynamical DNS server. Dynamical servers are particularly 
useful if your IP address may change over time, this is almost always the case for a home
computer that is linked to the internet via an ISP (Internet Service Provider). Later, 
once your website is at a mature stage more suitable domain names can be bought, 
for {\tt iCosmo} we used the common service {\tt http://www.123reg.com}.  

\subsection{Language Choice}

When developing a web interface, and an open source software package, the choice of a 
coding language is important. {\tt iCosmo} is written in IDL, which immediately created 
some obstacles to creating a web interface since IDL cannot be run directly from within 
a CGI script. Languages which are more conducive to web interface programming are 
C (which can be run directly within a browser), JAVA, PERL and python. One option 
for {\tt iCosmo} was therefore to re-code the entire package into something like JAVA or 
C, however this would have meant losing the features for which we chose IDL in the 
first place -- easy syntax, vast scientific libraries, easy plotting routines. 

When developing {\tt iCosmo} we found a way to call any programme from within a PERL CGI 
script. 
IPC::Open3\footnote{{\tt http://www.perl.com/doc/manual/html/lib/IPC/Open3.html}} 
(or Open2) creates a ``child'' process from the main PERL script which then runs 
(on the command line) and can return values to the PERL script. 
An example is given below: 
\\
\\
\noindent {\tt \#!/usr/bin/perl\\
print<<HTML;\\
<HTML>\\
<head>\\
<title>Hello World</title>\\
</head>\\
<body>\\
<font color="yellow">\\
 Hello World\\
</font>\\
HTML\\
\#this is a comment\\
\#do a simple calculation in PERL\\
\$x=2*2+2*2*2;\\
\#display the calculation on the webpage\\
print<<HTML;\\
2<sup>2</sup>+2<sup>3</sup>=\$x\\
HTML\\
\\
use IPC::Open3;\\
\$ENV\{'PATH'\} = '/Applications/itt/idl70/bin/';\\
\$progName = "idl";\\
open3(WRITEIDL,READIDL,ERRORIDL, \$progName);\\
print WRITEIDL "x=2\verb ^ 2+2\verb ^ 3+2\verb ^ 4 $\backslash$n";\\
print WRITEIDL "print,x $\backslash$n";\\
\$y =<READIDL>;\\
print<<HTML;\\
2<sup>2</sup>+2<sup>3</sup>+2<sup>4</sup>=\$y\\
HTML\\
\#end the HTML\\
print<<HTML;\\
</body>\\
</HTML>\\
HTML\\
close(WRITEIDL);\\
close(READIDL);\\
close(ERRORIDL);\\
exit;\\
}\\ 
\\
\\
In this example an IDL script is initialised by the Open3 command, when the command {\tt 
WRITEIDL} is called (which is named as the input to the IDL) actual IDL command 
lines can be entered which can then be read back into the PERL script using 
{\tt READIDL}. The result can then be printed into an HTML page in the usual 
way. Note that any code package could be run in this way, we use IDL as an example.

You may want to run some code on a machine that is not the server. For example 
this may occur if 
you cannot install some particular library or if more computing power is needed. The 
IPC::Open command can 
be used so that when a particular CGI script is run the server can {\tt ssh} 
into a machine that has the required software, perform the calculation on that 
machine and then output the results to the webpage. This is done by stacking 
the commands on the \$progName line e.g. 
{\tt \$progName = "ssh server@machine 'cd WebServer/CGI-Executables; command'"} 
which connects to a machine under the username server, changes directory to 
the CGI folder and then runs some {\tt command}. 
Most servers should come with some common software already installed, although installing 
some complex astronomical software may be difficult.

\subsection{Input and Output}
\label{Input and Output}
We have now outlined how to create a simple webpage and a CGI script that can run a 
particular cosmology code and output results to a webpage in a text format. Here will 
will discuss how inputs can be taken from a webpage and put into a CGI script. 

The key to web interactivity is the HTML \emph{form}. Forms allow a web user to input 
information into a webpage which can then be read by the server. You will find forms on 
almost every web page (from online shops to airline bookings). 
A form is basically a set of inputs and a `submit' button (forms can also be 
automatically submitted -- using javascript) 
when a form is submitted a new script is run which is given 
the variables input as arguments. The following example would create some boxes in which 
numbers could be input and then sent to the new CGI script. 
\\
\\
\noindent{\tt 
<form action="cgi-bin/newscript.cgi" >\\
<input type="text" name="omega\_m" size="5" value=0.3>\\
<input type="text" name="omega\_de" size="5" value=0.7>\\
<input name="submit" type="submit" value="Begin Calculations">\\
</form>
}
\\
\\
The variables {\tt omega\_m} and {\tt omega\_de} are submitted to {\tt newscript.cgi} 
(the variables can be seen in the URL delimited by \&'s or 
can be hidden using METHOD=``POST'' in the first line). There are many types of form 
input the most common are
\begin{itemize}
\item text : input a text value, numerical or alphabetical
\item checkbox : a tick-box which can take one of two values depending on whether 
it is selected or not
\item radio : an exclusive `or' i.e. value can take a number of different values
\item select : a drop-down box is created which gives a number of exclusive options
\end{itemize}
There is extensive online help for all these types of form input. 

In {\tt iCosmo} to read form data into a CGI script we used the {\tt ReadParse} PERL 
command which works in the following way 
\\
\\
\noindent {\tt \&ReadParse(*input);\\
\$omm = \$input{'omega\_m'};\\
\$omv   = \$input{'omega\_de'};\\
}
\\
\\
which would read in the submitted variables from above into the new PERL script as the 
variables {\tt \$omm} and {\tt \$omv}.

To create plots using PERL we used the common PGPLOT\footnote{{\tt 
www.astro.caltech.edu/$\sim$tjp/pgplot/}} software (we installed this for use in PERL using 
the excellent astronomical suite of tools provided by SciKARL\footnote{
{\tt http://astronomy.swin.edu.au/$\sim$karl/SciKarl}}). Within PERL a plot 
can easily be created using the following syntax or similar 
\\
\\
\noindent {\tt \#Create a graph:\\
\$ENV\{'PGPLOT\_PS\_BBOX'\} = "MAX";\\
pgbegin(0,"icosmo\_plot.gif/gif",1,1);\\
pubplot(\$xmin,\$xmax,\$ymin,\$ymax,0);
\\
\#black-on-white plots:\\
pgscr(0, 1.0, 1.0, 1.0);\\
pgscr(1, 0.0, 0.0, 0.0);\\
\\
pgsci(\$linecolour);\\
pgline(\$number, @x, @y);\\
pgend;\\
}
\\
\\
This would create a gif showing the data in the arrays {\tt @x} and {\tt @y}. 
To display this on a webpage the HTML {\tt img} command can be used
\\
\\
{\tt 
<img src="icosmo\_plot.gif" width="500px" alt=""/>}
\\
\\

To create a downloadable text file the following PERL syntax would write the structures 
{\tt \$x} and {\tt \$y} to a file 
\\
\\
{\tt
\$file="file.txt";\\
open (MYFILE, \$file);\\
for(\$start=1;\$start<\$number;\$start++)\{\\
\\
\$valueF=\$x[\$start];\\
printf MYFILE "\$valueF";\\
\$valueF=\$y[\$start];\\
printf MYFILE "\$valueF $\backslash$n";\\
\}
}
\\
\\
The following HTML would create a link to {\tt file.txt} 
\\
\\
{\tt
<a href='file.txt'> Download Text Output</a>}
\\
\\

\subsection{Browsers and Styling}
When creating a web interface one must always be careful with respect to 
browser compatibility. 
Issues can arise from even simple things such as choosing a font such that it will 
appear the same in each 
browser (we found this website particularly useful {\tt http://dustinbrewer.com/
fonts-on-the-web-and-a-list-of-web-safe-fonts/}, there are only a handful of 
fonts available) to an entire page appearing incorrectly. 

Incompatibility issues are usually created when obsolete HTML methods are used, 
this means that 
some new browsers will not recognise certain HTML commands and will render 
them in a different way 
than intended. The most up-to-date (c. 2008) standard for writing browser-proof HTML is 
to use \emph{styles}.

Styles are essentially small scripts that can be written separately (in the header) or 
within certain basic HTML commands that tell the browser how to render the object in 
question 
- that could be some text, a table, an image etc.  An simple example is given below
\\
\\
\noindent {\tt 
<a href="http://www.icosmo.org"><font color="\#66000">iCosmo</font></a>\\
\\
<a style="color: \#660000" href="http://www.icosmo.org">iCosmo</a>}\\
\\
\\
These two lines ostensibly do the same thing (create a red coloured hyperlink 
to {\tt iCosmo}) but 
the first line does not use styles and as such may appear incorrect in some browsers. 

Other problems can occur if, for example, a tables width is defined in \%, 
which some browsers recognise, whereas it should be defined in terms of pixels (px). 
When writing a web interface you should constantly check that whatever changes are 
made are accurately reproduced in all browsers. This can be done manually by 
installing some of the most common browsers (Firefox, Opera, Safari, Internet Explorer) 
or by using an online browser checker - we found this site particularly useful {\tt 
http://browsershots.org/}.

\subsection{Version Tracking}
When creating a web interface keeping track of different versions, and what
changes have been made is
important so that the site can be restored to any point in its creation if 
things go wrong. There 
are many commercially available and open source ways which can back up websites 
and code, the most commonly used is SVN (SubVersioN). 

For {\tt iCosmo}, creating a web site in collaboration, we found that 
using a self-made wiki page 
was the optimal solution. We used the public wiki pages 
{\tt http://www.pbwiki.com} and on our 
{\tt iCosmo} wiki page created a ``Blockbuster'' style table. 
If one person wanted to edit the 
site they ``checked out'' the code (downloaded as a zipped file), 
and made a note on the wiki. They 
would then ``check in'' and upload a new version. All versions are 
stored on the wiki page and this 
prevented multiple people working on the same code simultaneously as well as saving the 
old versions for backup.  

\subsection{Help Solutions and Documentation}
When creating a web interface for cosmology, and following the `transparent box' 
approach that we have taken with {\tt iCosmo}, 
sufficient material should be provided and links to relevant publications made 
(at the very least links to other sites that contain good material should be added). 

For {\tt iCosmo} we have made a multi-layered suite of help resources. 
The simplest is to add help 
directly as text to the webpages in question. Links can easily be made 
to publications available 
on online archives. The next layer is to create more information on a 
different page and make this 
an evolving and interactive document. The most common forms of these 
interactive documents are wiki 
pages and online word processors (e.g. GoogleDocs). For {\tt iCosmo} we have used 
{\tt http://www.pbwiki.com} to create a number of wiki pages that are 
edited and updated by a group
of experts in each field we have covered. The GREAT08 challenge (Bridle et al., 2008; 
{\tt http://www.great08challenge.info})  has made good use of 
GoogleDocs to provide online documentation. 

The next layer of interactivity is to provide an online forum for users of your site to 
discuss the
science, results and the webpage. In cosmology we are fortunate to have 
{\tt http://www.cosmocoffee.info}. For {\tt iCosmo} we have provided a discussion forum 
hosted by cosmocoffee. 

To make posting a website onto network sites (e.g. Facebook, MySpace) 
and bookmark sites (e.g. Twitter, Digg) easy for a user it is common to 
add small links at the bottom of a page that will automatically post to 
these websites. For {\tt iCosmo} we used the free service 
provided by {\tt http://www.sharethis.com} which provides a snippet of 
HTML that once pasted into a website compresses all the posting links into one button.

Finally video (or audio) tutorials can be provided in addition to text based and 
interactive help. Video 
tutorials can allow the web site to be demonstrated to the user and for 
all the aspects of the site 
to be presented. For {\tt iCosmo} we choose YouTube to host the videos 
online - YouTube also 
allows videos to easily be embedded directly into HTML. Alternatively 
you can host you own videos, or publish audio as a pod-cast.  
To record on-screen video and audio 
there are many cheap software options available; we chose {\tt iShowU} for Mac. 

\subsection{A Worked Example}
As a worked example we have provided some simple templates of an 
HTML page and a CGI page in Appendix B. Working versions of these can be found on 
\begin{itemize}
\item {\tt http://www.icosmo.org/WebPaper.html}
\item {\tt http://www.icosmo.org/cgi-bin/WebPaper.cgi}
\end{itemize}
In these examples we show how each of the components described in this Section can be 
brought together to create a simple cosmology interface. We show how styles and forms are 
used as well as some other HTML tricks, for example inserting an image, creating Greek 
letters and making lists. In the CGI example we take some data from the HTML page and 
use this data to perform a calculation using the iCosmo IDL software.

We have outlined how to create a web interface for cosmology calculations from scratch. 
This can be done by through the following 3 steps:
\begin{itemize}
\item
Create a static HTML site to collect user input. This can be done using the forms discussion in Section \ref{HTML} or the template provided {\tt http://www.icosmo.org/WebPaper.html}.
\item
Write a CGI script to run your program (e.g. an IDL .pro file). This is discussed in Section \ref{Input and Output}, 
and an example can be found at {\tt http://www.icosmo.org/cgi-bin/WebPaper.cgi}.
\item
Create an interface for displaying the results. See {\tt http://www.icosmo.org/cgi-bin/WebPaper.cgi} for an example
\end{itemize}

Some users may wish to simply extend the capability of an existing site, such as 
{\tt iCosmo}, this can be done by using the templates provided. 
The {\tt iCosmo} website source code will be released in {\tt iCosmo} v1.2 
({\tt http://icosmo.pbwiki.com}). 

\section{Online Cosmology}
\label{Online Cosmology}
The internet has been harnessed in astronomy and cosmology 
as a communication tool and as a source of processing 
power for individual calculations. The move towards using the 
interactivity of the internet 
advocated here is similar in motive and result to \emph{cloud computing} where 
computational processes are not performed locally but within the cloud (remote, 
anonymous, machines on the internet). 
We refer to the use of the internet in this fashion, for cosmological 
calculations, as \emph{cloud cosmology}.

It may appear strange to an expert in web design or cloud computing 
that we are using the relatively well known CGI approach to writing 
webpages as an example of cloud computing. 
However the definition of 
cloud computing does not stipulate the method by which the cloud is contacted. 
We advocate the use of CGI scripts because of their 
simplicity, easy-of-use and the vast amount of 
scientific libraries that are available for them.  

Other recent suggestions for using the interactivity of the internet to increase 
productivity in astronomy and cosmology are 3D environments and the virtual observatory. 
3D environments (e.g. Hut, 2007) can be used for outreach activities and as 
visualisation tools for complex data sets. The concept of the 
virtual observatory (e.g. Szalay \& Brunner, 1999) 
is that astronomers can access vast libraries of  
pre-observed data, where any desired observation has a high probability of 
already existing, thus making the need for a new observation redundant. 

We have outlined the particular solutions that we have found that will enable  
cosmologists to turn their code
into a web interface. Looking ahead 
we envisage a move to a more open culture of software dissemination where 
not only articles are published but the code used to create the science in the 
article is also 
made public. Here we define a number of tiers possible of involvement 
\begin{itemize}
\item As a minimum one can publish the code, related to an article, online.
\item 
Secondly the code can be published as part of a coherent software package 
e.g. {\tt iCosmo}. 
In this case, in order to become integrated into the package both the code and the 
science need to be refereed to some degree to ensure consistency
\item 
Thirdly the code can be published and turned it into a web interface, this could be as an 
extension of an existing interface (e.g. {\tt http://www.icosmo.org}) or as a stand alone interface.
\end{itemize}

The {\tt iCosmo} source code itself (Refregier et al., 2008) 
is an example of the third tier, 
both open source and web interactivity. 
Kitching et al. (2008) (a cosmological systematics investigation) is an example of 
the second tier, some source code which has been integrated into an existing 
software package as an additional module. 

{\tt iCosmo} can be used as a platform for all three tiers. The source code is very 
modular 
and can easily be extended, and because {\tt iCosmo} is an existing software package new 
add-on modules can be published in each new release. Finally the methods 
outlined in this article show how an {\tt iCosmo} module can be turned into an 
interactive webpage. New interactive webpages could either be entirely independent or 
could be incorporated in the main site. 

\section{Conclusion}
\label{Conclusion}
Cosmology had made good use of the internet so far as either a public 
repository for articles and code, or as a large-scale computational farm. 
However the interactive 
potential of the web, that can allow cosmologists to perform complex 
calculations online, has thus far been largely overlooked. 

We envisage a move towards cloud cosmology in which computationally demanding 
and complex processes can be performed non-locally. 

In this article we have outlined the some simple processes 
that we have learnt in the creation of 
an online, interactive, low-redshift (dark Universe) cosmology calculator. 
We have outlined 
how a simple web page is created, and hosted, and shown how 
cosmological code can be turned into an interactive web site. 
Throughout we have used the 
example of the interactive web site {\tt http://www.icosmo.org} 
which is powered by the open source 
software {\tt iCosmo} -- for a description of the interactive 
features on this site see Appendix A.

Finally we described three levels integration that could be 
adopted by a cosmologist wanting to make their code easily accessible online. 

\section*{Acknowledgments}
TDK is supported by the Science and Technology Facilities Council,
research grant number E001114. We thank all the {\tt iCosmo} contributors 
Filipe Abdalla, David Bacon, Sarah Bridle, Alain Gueguen, 
Benjamin Joachimi, Daniel Kubas, Justin Read, 
Shaun Thomas, Jochen Weller as well as Luca Amendola, Nabila Aghanim, 
Anna Cibinel, Marian Douspis, Robert Feldman, Oliver Hahn, 
Alan Heavens, Ofer Lahav, Andy Taylor. 
For help in comparing Fisher matrix calculations we also thank  
Jiayu Tang, Peter Schneider, 
Martin Kilbinger, Jan Hartlap, Martin Kunz and Fransico Castander. 
For help in hosting the website we thank ETH Zurich and 
Axel Beckert for invaluable and expert help.


\appendix

\section{User Guide for http://www.icosmo.org}

In this Appendix we describe the interactive features of {\tt iCosmo}, the products 
available from the website and how to access the main features of the web interface. 

\subsection{Beginning Calculations}
\label{Beginning Calculations}

The {\tt iCosmo} interactive website 
allows you to calculate cosmological functions, cosmology observables and 
Fisher matrices. These are summarised in Table \ref{summary}. We always calculate the 
cosmology functions. The three different calculation scopes will be described 
in more detail in Section \ref{Interactive Pages}.

\begin{table*}
\begin{center}
\begin{tabular}{|l|c|}
\hline
 Scope & Products \\
\hline
Cosmology Functions & Distances,$H(z)$,Growth Factor\\
                    & Linear Power Spectrum, Non-Linear Power Spectrum \\
\hline
Cosmology Observables & Lensing Power Spectrum, Baryon Distance Scales\\
                    & Supernovae Type-Ia Magnitude Distance Relation\\
\hline
Cosmology Errors & Lensing Fisher Matrix\\
                 & BAO Fisher Matrix, Supernovae Type-Ia Fisher Matrix\\
\hline
\end{tabular}
\caption{The range of products available on the website. Cosmology Functions are always 
calculated. The user can investigate the effect of changing cosmological parameters 
in the case of Cosmology Functions and Cosmology Observables.}
\label{summary}
\end{center}
\end{table*}

Throughout all calculations we use the following set of cosmological parameters $\Omega_m$, $\Omega_{\rm DE}$, $w_0$, $w_a$, $\sigma_8$, $\Omega_B$, $h$ and the spectral index $n_s$. 
The curvature density $\Omega_K$ is determined through the choice of $\Omega_m$ and $\Omega_{\rm DE}$ where $\Omega_K=1-\Omega_m-\Omega_{\rm DE}$ at all times.

There are two ways in which the interactive part of the {\tt iCosmo} web site can be accessed. Either though the QuickStart panel on the front page or through a step-by-step process that allows the 
user to customise the exact scope and details of calculations required.

\subsection{QuickStart} 
\begin{figure}
\resizebox{84mm}{!}{
\includegraphics{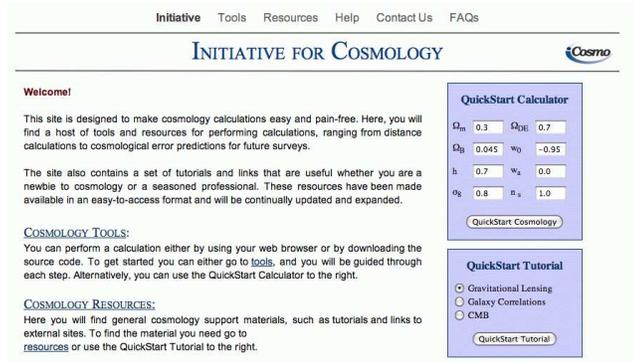}}
\caption{The Front Page of the {\tt iCosmo} web site. The QuickStart Calculator panel can be see on the right hand side. To access the stepwise interactive pages click on the ``Cosmology Tools'' link or on the ``Tools'' link in the navigation bar across the top of the page.}
\label{FrontPage}
\end{figure}

You will find the QuickStart panel on the front page of the {\tt iCosmo} website, see Figure \ref{FrontPage}. To start simply 
fill in the values of the cosmological parameters and click on ``QuickStart Cosmology''. You will be taken to a waiting page whilst the calculations are performed. When the calculations are completed you will be redirected to the Cosmology Functions interactive tab described in Section \ref{Interactive Pages}.

\subsection{Step-by-Step}
\label{Step-by-Step}
\begin{figure}
\resizebox{84mm}{!}{
\includegraphics{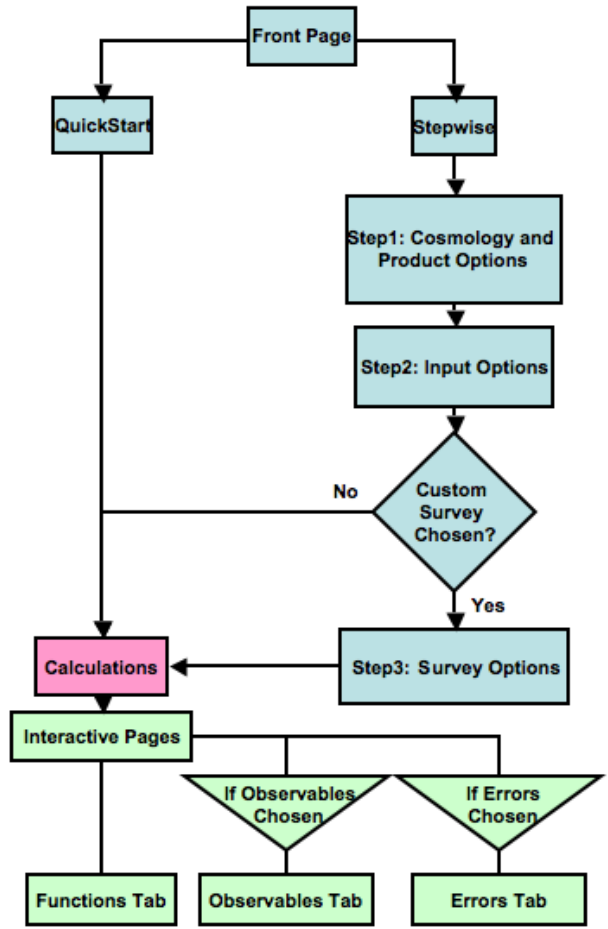}}
\caption{The outline of the website in the form of a flow diagram highlighting the inter-relations between the optional inputs and the resultant outputs. A diamond represents a logical `for' i.e. for option A do this or for option B do that. A triangle represents a logical `if' i.e. if C is chosen then do this. }
\label{StepFlow}
\end{figure}

To start your calculation and to choose Cosmology Observables and Errors 
options from the beginning the user can begin a stepwise customisation of the calculations needed. This is reached by clicking on any of the links to ``Tools'' on the front page, Figure \ref{FrontPage}, and then clicking on ``Interactive Web Tool''. Throughout the stepwise process the user will only be shown options relevant to the calculations requested, for example if the lensing power spectrum or lensing Fisher matrices are not required the user will never see any lensing-related options. 
The structure of this flow is outlined in Figure \ref{StepFlow}.

{\bf Step 1} allows the user to specify the cosmology, or cosmologies (see Section \ref{Multiple Cosmologies}) that they wish to use in the calculation. In this step the range of products that will be calculated is also chosen, the products are arranged into the three scopes detailed in 
Table \ref{summary}. To advance to the next stage the user clicks on ``Begin Calculations''.

{\bf Step 2} will show some further options relevant to the products chosen in Step 1. The baseline options for the Cosmology Functions are the minimum and maximum redshift for the distance, Hubble parameter and growth factor calculations. For the matter power spectrum the minimum and maximum k-range as well as the functional form used in the calculations can be chosen. The types of option available are shown in Table \ref{step2options}. For the matter power spectrum the linear form can be calculated using either Eisenstien \& Hu (1997) (with or without baryon oscillations) or Bardeen et al. (1986). The non-linear power spectrum can be calculated using either Peacock \& Dodds (1996) or Smith et al.(2003). 

If either Cosmology Observables or Cosmology Errors were chosen in Step 1 then the choice of survey will be given in Step 2. If ``Make Custom'' is chosen then the user will be taken to a further Step where the survey parameters relevant to the type of probe required can be customised. Alternatively a pre-defined survey can be chosen, the available pre-defined options are detailed in Section \ref{Surveys}.

{\bf Step 3} will appear if ``Make Custom'' is selected in Step 2. In Step 3 the user will be 
shown survey parameters specific to the type of probe that is required, either lensing, baryon acoustic oscillations (BAO) or type-Ia supernovae. The details of the options available for each of the probes are:
\\

\noindent Lensing: 
\begin{itemize}
 \item The area of the survey in square degrees
 \item The number $n$ of tomographic redshift bins used to construct lensing power spectra, $n(n+1)/2$ spectra will be calculated. Note that the bin boundaries will be chosen such that the number of galaxies in each bin is approximately constant
 \item The surface number density of galaxies in galaxies per square arcminute that can be used for a cosmic shear analysis
 \item The intrinsic variance in the measured ellipticity of galaxies, here we use the convention that the variance in measured shear is $\sigma_{\epsilon}=\sigma_{\gamma}/2$ as explained in Bartelmann \& Schnieder (2001)
 \item The uncertainty $\delta_z(z)$ in the redshift of each galaxy, using the common $(1+z)$ scaling $\sigma_z(z)=\delta_z(z)(1+z)$. A typical broad band photometric redshift survey should have $\sigma_z(z)/(1+z)\sim 0.02$ to $0.05$ whereas a spectroscopic survey may have $\sigma_z(z)/(1+z) < 0.001$.
 \item The median redshift for the number density distribution of galaxies. We employ the commonly used Smail et al. (1994) formula
\end{itemize}

\noindent Supernovae:
\begin{itemize}
 \item The area of the survey in square degrees that is multiply imaged to allow supernova studies
 \item The surface number density of supernovae in supernovae per square arcminute. Note that this depends on the total time of the survey, the depth of the survey (magnitude limit) and the intrinsic efficiency of a galaxy in producing supernovae
 \item The intrinsic variance in supernovae apparent magnitude, this is typically $\sigma_m\sim 0.15$.
 \item The uncertainty $\delta_m$ in the measured apparent magnitude of a given supernovae, this value will depend on the particular instrument that is used in the observations. For a space-based instrument $\delta_m\sim 0.02$ and for a ground-based instrument the uncertainty may be $\delta_m\sim 0.05$.
 \item The minimum and maximum redshifts for the number density of supernovae. Note that for simplicity we will assume that the number of supernovae in each redshift bin is constant in redshift. To calculate the number of redshift bins we set the shot noise uncertainty on the apparent magnitude in a given bin to be $1/1000$ the intrinsic magnitude dispersion as is typically assumed Tegmark et al. (1998) and Huterer \& Turner (2001).
\end{itemize}

\noindent BAO:
\begin{itemize}
 \item The area of the survey in square degrees
 \item The number of redshift bins in which to measure the BAO scale. Note that the bin boundaries will be chosen such that the number of galaxies in each bin is approximately constant
 \item The median redshift for the number density distribution of galaxies. We employ the commonly used Smail et al. (1994) formula
\end{itemize}

\begin{table*}
\begin{center}
\begin{tabular}{|l|c|}
\hline
Product(s) & Options \\
\hline
Distances, Hubble Parameter, Growth Factor & $z_{\rm min}$ , $z_{\rm max}$\\
\hline
Matter Power Spectrum & $k_{\rm min}$, $k_{\rm max}$\\
                      & Linear: Eisenstien \& Hu (1997) (with or without baryon oscillations)\\
                      & or Bardeen et al. (1986)\\
                      & Non-Linear: Peacock \& Dodds (1996) or Smith et al.(2003)\\
\hline
Lensing Power Spectrum or Lensing Fisher Matrix & $\ell_{\rm min}$, $\ell_{\rm max}$, Survey\\
\hline
BAO Distance Measure or BAO Fisher Matrix& Survey\\
\hline
Supernovae Magnitude-Distance or Supernovae Fisher Matrix & Survey\\
\hline
\end{tabular}
\caption{The options given in Step 2 relevant for each of the products chosen in Step 1.}
\label{step2options}
\end{center}
\end{table*}

\begin{figure}
\resizebox{84mm}{!}{
\includegraphics{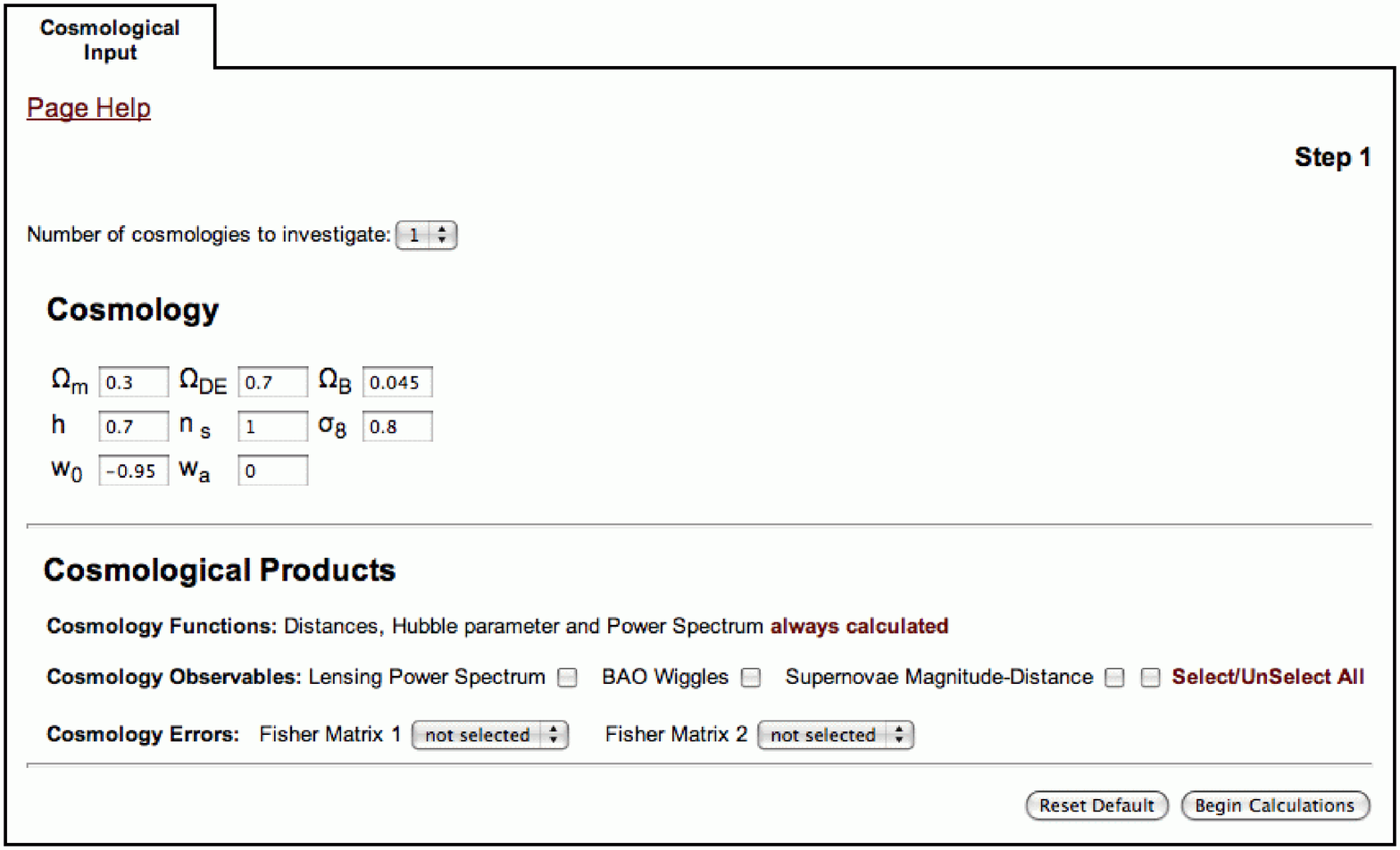}}
\caption{The first step in the stepwise process of specifying the inputs to the calculation.}
\label{StepWise}
\end{figure}

\subsection{Surveys}
\label{Surveys}
Both the observables and the Fisher matrix parameter error predictions depend on the design of the experiment. The user will find on Step 2 that for each lensing 
observable and Fisher matrix required we have provided the choice of either a custom survey or a 
current/future survey. The references for the surveys shown are:
\begin{itemize}
\item Dark Energy Survey (DES): DES Collab (2005)
\item Dark UNiverse Explorer (DUNE): Refregier et al. (2008b)
\item Large Synoptic Survey Telescope (LSST): Tyson et al. (2006), Ivezic et al. (2008)
\item Panoramic Survey Telescope \& Rapid Response System (Pan-STARRS 1): Kasier et al. (2004)
\item SuperNovae Acceleration Probe (SNAP) : SNAP Collab (2005), Huterer et al. (2005)
\end{itemize}
The suite of surveys available will continually be updated and any suggestions can be sent to {\tt 
help@icosmo.org}

\subsection{Multiple Cosmologies}
\label{Multiple Cosmologies}
In Step 1 the user can choose between one or two cosmologies. This feature is to 
enable the sensitivity of cosmological functions and observables to certain cosmological 
parameters (or combination or parameters) to be investigated. By selecting two cosmologies from the drop down button on Step 1 the user can fill in the values of each cosmological parameter. On the interactive pages the user will then see two lines, one for each cosmology, above either the cosmological functions or observables being plotted.

After the last Step (either Step 2 or Step 3 depending on the products and level of customisation needed) the user is taken to a waiting page whilst the calculations are performed. Note that there are certain `illegal' values of the cosmology parameters and customisable variables such as $\Omega_m < 0$ or $k_{\rm min} > k_{\rm max}$ that the {\tt iCosmo} source code will not accept. If an illegal value is passed to the interactive calculator the waiting page will display a message detailing the values that are discrepant and display a button with which the user can return to the previous step. After the calculations the user will be redirected to the Interactive results pages described in the next Section \ref{Interactive Pages}.

\subsection{Interactive Pages}
\label{Interactive Pages}
In this Section we will describe the Interactive results web pages. When the user arrives at the interactive results page they will be presented with a number of \emph{tabs} one can change between any of the tabs simply by clicking on the relevant headers. The Cosmology 
Functions tab will always be present, since these are always calculated, in addition there may also be a Cosmology Observables and a Cosmology Errors tab. 
The user can click on any tab and change between tabs in any order or combination. In addition the 
navigation bar at the top of the page allows the user to link to other {\tt iCosmo} pages. 

The options that are given 
on each tab are shown in Table \ref{taboptions}, where an ``\&'' means that each option can be shown simultaneously and an ``OR'' means that the options are mutually exclusive.

\begin{figure}
\resizebox{84mm}{!}{
\includegraphics{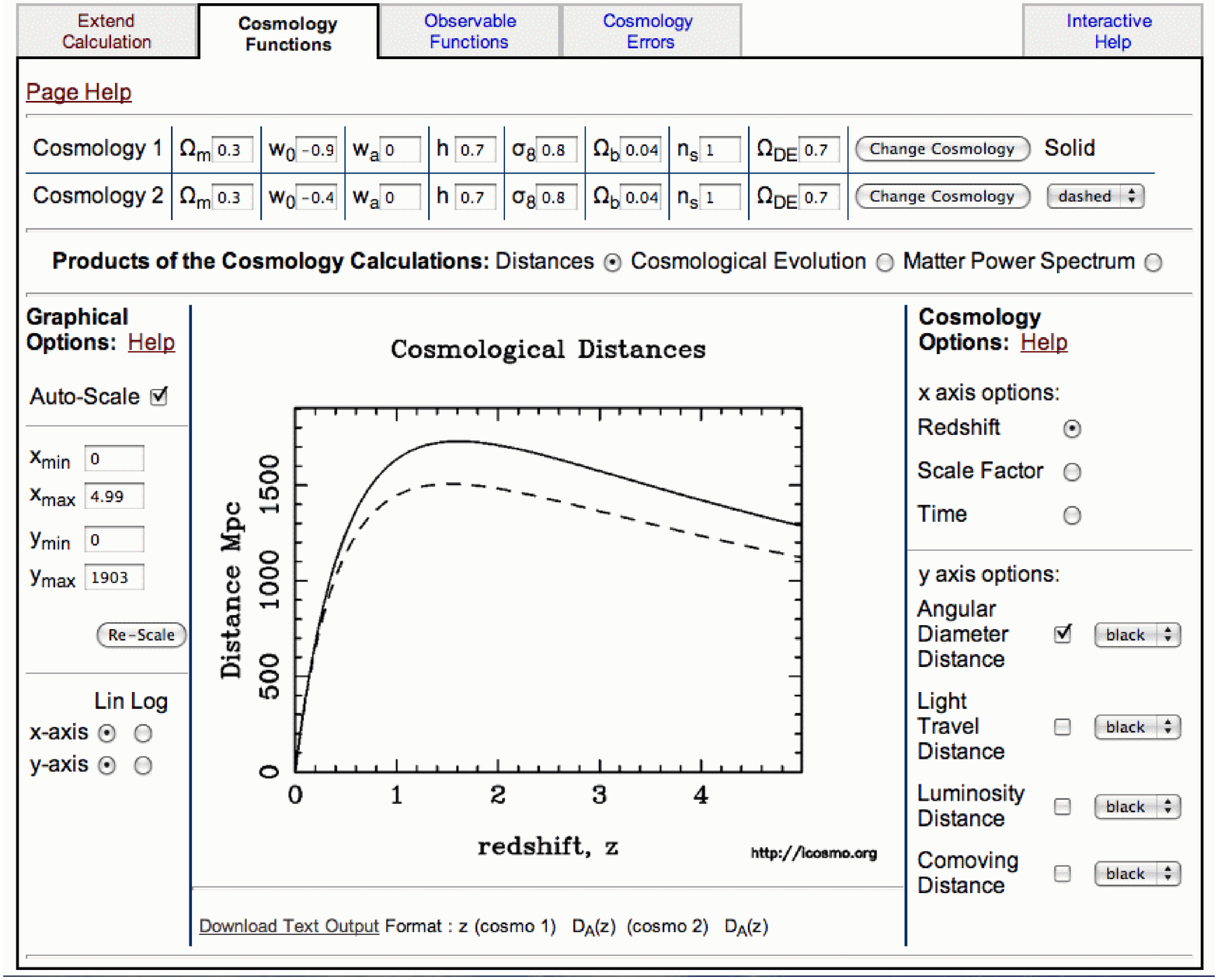}}
\caption{The Interactive results tabs in {\tt iCosmo} web site. 
Here we show the Cosmology Functions tab, 
in this example we have chosen two cosmologies and the distances button is shown.}
\label{Tabs}
\end{figure}

The results tabs are all arranged using a common template which be seen in Figure \ref{Tabs}. 
This template is split into a number of regions. 
The top region is dedicated to the input cosmology used, the survey parameters (in the case of the Cosmology Observables and Errors tabs) and the options available for the particular tab. 

Associated with each cosmological parameter is a box containing the current fiducial value. To change the fiducial value the user can simply enter an alternative value and click on ``Change Cosmology''. The user will then be redirected to calculation page where the cosmological products will be re-calculated using the new cosmology.

The central part of each tab is the graphical window. When any of the options in a tab are changed the plot will be \emph{automatically updated}. The plot is provided on screen in gif format, to save the plot as a gif most browsers provide a save option when an image is right-clicked. In addition to the gif format we provide an encapsulated postscript version of every plot, this can be downloaded by clicking on the plot. 

In every tab, to the left of the plot, there are the {\bf graphical options} these give the user the ability to control the x and y-axis ranges of the plot using the ``Re-Scale'' button. By default the autoscale button is selected. When autoscale is selected the plot will be automatically scaled such that the function plotted will fit within the viewport. We recommend that the autoscale function is enabled in most situations. In addition to rescaling the graph the type of axis can also be changed -- from linear to logarithmic (base $10$) or vice versa -- by clicking on the relevant buttons.

Below the plot, in each tab, is a ``Download Text Output'' link. By clicking on this link the user can download the data plotted at the present time in an ascii table format for later use. The format of the ascii file is shown to the right of the link and is updated so that it is always relevant to the particular function being plotted.

\begin{table*}
\begin{center}
\begin{tabular}{|l|c|c|c|}
\hline
{\bf Tab} & {\bf Button} & \multicolumn{2}{|c|}{{\bf Options}}\\
\hline
    &        & {\bf x-axis} & {\bf y-axis} \\
\hline
Cosmology Functions & Distances & redshift, & Angular Diameter Distance\\
&&scale factor,& \& Luminosity Distance \\ 
&&look-back time& \& Light Travel Distance \\
&&& \& Comoving Distance \\
\hline
& Evolution & redshift,  & Hubble Parameter \\
&&scale factor,& OR Growth Factor \\
&&look-back time& OR Angular Size\\
\hline
& Matter Power Spectrum &      wavenumber, $k$                      & Linear \\
&&& \& Non-Linear, Redshift \\
\hline
Cosmology Observables & Lensing Power Spectrum &  azimuthal wavenumber, $\ell$   & Cross or auto correlation spectra,\\
&&& Survey Error Bars\\
\hline
& Baryon Oscillation Distance Scale &        redshift                 & Radial \\
&&& OR Tangential modes, \\
&&& Survey Error Bars\\
\hline
& Supernovae Magnitude-Distance Relation &      redshift                   &  Survey Error Bars\\
\hline
& & \multicolumn{2}{|c|}{{\bf Options}}\\
\hline
Cosmology Errors & Fisher Matrices & \multicolumn{2}{|c|}{Cosmological Parameters. Display Fisher 1 \& Fisher 2 \& Combined} \\
\hline
\end{tabular}
\caption{The options for each tab and button on the interactive webpages.}
\label{taboptions}
\end{center}
\end{table*}

To the right of the plot in each tab are the options that are relevant to the particular products that are being plotted, we will now explain each tab in turn. 

\subsubsection{Cosmology Functions Tab}
The cosmology functions tab presents the calculations of the basic cosmology functions, and 
will always be present in the interactive section. 
To change between the functions the user can click on the relevant button. When the button 
is changed the ``Cosmology Options'' on the righthand side of the graph will change to reflect the new options available to the user. 
\\
 
\noindent {\bf Distances} When the distances button is selected the angular diameter, luminosity, comoving and light travel distances can be plotted. Any combination 
of these distances can be plotted by selecting the relevant checkboxes -- 
colours for each function can 
also be chosen independently using the drop down menu. These functions can be plotted as a 
function of: redshift $z$, scale factor $1/(1+z)$ or look-back time. 
\\

\noindent {\bf Evolution: Hubble Parameter, Growth, Angular Size}
When the cosmological evolution button is selected the Cosmology Options will change to 
allow the growth factor, Hubble parameter and angular size of galaxies to be plotted. 
The user can change the graphical output by selecting the relevant button, and change the 
colour of the line accordingly. All of these functions can be plotted as a function of redshift, 
scale factor or look-back time. 
The angular size is simply a modification of the angular diameter distance, we included it in this 
button (cosmological evolution) since it is not strictly a distance measure but a derived quantity 
that changes as a function of redshift. It is also a quantity that is of use to for some 
observational proposals. The growth factor is found by solving the growth factor differential 
equations. 
\\

\noindent {\bf Matter Power Spectrum}
By clicking on the matter power spectrum button the page will be updated with options 
relevant to the matter power spectrum. If the QuickStart button was used then the linear power 
spectrum will have been calculated using the Eisenstein \& Hu (1997) formulae (with baryon 
oscillations included), and the non-linear correction using Smith et al. (2003). The choice of 
linear and non-linear power spectrum (Table \ref{step2options}, Step 2) will be shown in the bottom righthand corner of the tab. Both the linear and non-linear power spectrum can be plotted, 
in addition the redshift of the power 
spectrum can be chosen from a grid of $50$ finely spaced redshifts between $0\leq z< 5$. 

\subsubsection{Cosmology Observables Tab}
\label{Cosmology Observables Tab}
The Cosmology Observables tab will appear if some observables have been selected for calculation. 
Below the standard cosmology parameters will appear buttons that match to each of the observables 
requested. For every observable the $1$-$\sigma$ error bars for the survey chosen can be shown 
(or suppressed) at any time by selecting (de-selecting) the {\bf show errors} checkbox. Naturally 
for any of the displayed functions the colour of the line drawn is always given as an option.
\\

\noindent {\bf Lensing Power Spectrum} The lensing power spectrum in fact 
consists of multiple individual spectra, for $n$ tomographic bins there will be $n(n+1)/2$ spectra. 
These are the auto-correlation of the cosmic shear 
signal in each redshift bin, and the cross correlation of the cosmic shear signal between redshift
bins. The auto-correlation for the lowest redshift bin is initially shown, the drop down box 
labelled ``$C_{\ell}$ Bins'' allows the user to plot any of the spectra calculated for the 
cosmic shear survey (the number available will depend on the number of redshift bins chosen).
\\ 

\noindent {\bf Baryon Wiggles} The baryons oscillation distance scale can be measured in either the
radial direction (parallel to the line-of-sight) or in the tangential direction (perpendicular 
to the line-of-sight). The user can show either of these distance measures using the Cosmology 
Options buttons. For a recent discussion of baryon oscillations see Rassat et al. (2008).
\\

\noindent {\bf Supernovae Magnitude-Distance} The supernovae magnitude-distance option shows the 
predicted apparent magnitudes of Type-Ia supernovae as a function of redshift.

\subsubsection{Cosmology Errors Tab}
\label{Cosmology Errors Tab}
The cosmology errors tab will appear if Fisher matrices have been requested in Step 1 (see 
Section \ref{Step-by-Step}). Fisher matrices calculate the expected marginal errors for a given 
cosmological probe and survey design. The cosmology errors tab will show either one or two 
sets of errors depending on the number of Fisher matrices requested. 
The $1$-$\sigma$ marginal errors are shown under each cosmological parameter at the top of the tab,
these take into account all the degeneracies that exists between all of the parameters shown.

Some cosmological probes are only sensitive to particular parameters, in the case that neither 
Fisher matrix can constrain a parameter it will not be shown. In the case that the first 
Fisher matrix is sensitive to more parameters than the second then any parameters to 
which the second is insensitive will be given a ``--'' in place of a marginal error. We also 
show the dark energy equation of state pivot redshift error $w_p$, this is the minimum error on 
function $w(z)$ which we parameterise using $w_0$ and $w_a$ (Chevallier \& Polarski, 2001; 
Linder, 2003). We also show the Dark Energy Task Force (Albrecht et al., 2006) Figure of 
Merit (FoM) which is proportional to the 
reciprocal of the area constrained by the $w_p$ and $w_a$ ellipse FoM$=1/(\Delta w_a\Delta w_p)$.

A prior represents the extra (or \emph{a priori}) knowledge that one may have on a 
particular parameter, we allow the user to add a Gaussian prior to any parameter. 
To add a prior to any parameter a user can simply fill in the boxes under the parameter in 
the ``Gaussian Prior'' row at the top of the tab and click on ``Add Prior''. The prior will be 
added to all Fisher matrices shown (one, two and combined), and the marginal errors and 
graphical window will be automatically updated. 

We graphically display the predicted marginal errors as contour plots. The graph can show the 
$1$-$\sigma$ marginal error contours for any of the parameters to which to methods are sensitive. 
The contours will always be ellipses since in all Fisher matrices we have assumed Gaussian 
distributed data (Tegmark et al., 1997). 
The user can chose between any parameter combination by using the drop down menus under 
Cosmology Options. In addition to changing the colour of the line drawn the ellipses can also be 
filled with a colour chosen from the drop down menus. To show a Fisher matrix (or suppress it) 
the check box to the left of the Fisher matrix description can be selected (or de-selected). 

If two Fisher matrices have been calculated the user will now have the further option of combining 
the constraints. This is done by selecting the check box to the left of ``Combined Fisher'' under 
the Cosmology Options. When the Fisher matrices are combined there will now be an additional 
row of marginal errors shown at the top of the tab and an additional constraint drawn in the 
graphical window. 

The Fisher matrices are calculated using the formalism outlined in following publications. For lensing we use Amara \& Refregier (2007), this calculation has been extensively tested and found to be 
in agreement with a number of independent codes, these include codes written by: Sarah Bridle, Filipe Abdalla, Jiayu Tang, Shaun Thomas, Peter Schneider, Martin Kilbinger, Jan Hartlap and Benjamin Joachimi as well as a 3D cosmic shear code (Heavens, 2003; Heavens et al. 2006). For the BAO 
calculation we use the formalism for spectroscopic surveys developed by 
Blake et al. (2006), this has been tested against, and found to be in agreement with 
independent codes written by Martin Kunz, Fransico Castander as well as the DETFfast website, for 
a description of these implementations see Rassat et al. (2008). For 
supernovae we use the Fisher matrix formalism described in Tegmark et al. (1998) and 
Huterer \& Turner (2001). 

\subsubsection{Extending the Calculation \& Help}
\label{Extend}
The final tabs on the interactive website allow the user to change the calculation from what they 
have already calculated and get some more Help. 
\\

{\bf Help} If any extra help is needed we provide some short pragmatic $\ls 1$ minute clips 
where particular aspects of the interactive tabbed environment are demonstrated. 
These are hosted on an external \emph{YouTube} channel {\tt http://www.youtube.com/icosmology}.
\\

{\bf Extend Calculations} On the extend calculations tab the user will find a copy of Step 1 
(see Section \ref{Step-by-Step}) but with the values and scope already calculated filled in 
with the currently selected products and variables. The 
user can then change the options in any way just like Step 1, either extending or simplifying 
the calculation or just changing some of parameters. By clicking on ``Extend 
Calculation'' the user is taken to Step 2. 

\subsection{Conclusion}

In this Appendix 
we have introduced the interactive web tool which is part of the \emph{Initiative for 
Cosmology}, a general cosmology resource. The interactive web tools are powered by {\tt iCosmo}, 
an easy-to-use open-source cosmology calculator presented in Refregier et al. (2008) and on 
{\tt http://icosmo.pbwiki.com}. In addition to the source code and interactive online tools 
the \emph{Initiative for Cosmology} provides a suite of online resources and tutorials.  
  
We have explained the structure and 
connectivity of the interactive pages, and described the functionality in a pedagogical fashion. 
This Appendix should also be relevant as a `user guide' for the website. 
The interactive website allows the real-time calculation of 
\begin{itemize}
\item cosmological distances
\item hubble parameter
\item growth factor
\item linear and non-linear matter power spectra
\item lensing power spectra
\item baryon distance scales
\item supernovae magnitude-distance relation
\item lensing Fisher matrices
\item baryon acoustic oscillation Fisher matrices
\item supernovae Fisher matrices
\end{itemize}
For any of these calculations the cosmology parameters, numerical ranges and survey design are  
all free variables that can be input directly in to the website. 

The interactive web tools that are presented here form part of a continually evolving and dynamic 
resource. We have presented v$1.0$ in this Appendix, the website will be updated with new versions 
on a regular basis. To discuss this website please visit the {\tt iCosmo} forum on 
{\tt http://www.cosmocoffee.info}, or email {\tt help@icosmo.org}.

\onecolumn
\section*{Appendix B : A Worked example} 

\noindent {\bf HTML Example.} Here is an HTML script that can be used to create the 
page {\tt http://www.icosmo.org/WebPaper.html}, we also display this HTML 
on the page itself.

\noindent {\tt
\begin{lstlisting}[numbers=left]
<html>
<head>

<title>A Simple Cosmology HTML Page</title>

<style type="text/css">
htest
{
font-family: arial;
font-size: 10px;
color: \#660000
}
</style>

</head>

<body>

<h1>Welcome!</h1>

<p>
/*@ This HTML page accompanies the paper <em>Cloud Cosmology : Building the Web Interface for iCosmo</em>. Here we will demonstrate some simple HTML syntax including forms, styles and images. We could not possibly give examples of all possible HTML syntax here, for extensive tutorials follow these links:@*/
</p>
<ul>
<li>(This is also an example of how to make a list and hyperlinks)</li>
<li><a href="http://www.w3schools.com">w3 schools</a></li>
<li><a href="http://www.htmlcodetutorial.com/">html tutorial</a></li>
</ul>

/*@<p>The HTML used for this page is shown at the bottom of this page, or can be viewed by using the <em>show source code</em> option in most browsers.</p>@*/

<h2>Styles</h2>

<p>There are two ways in which the style of text can be changed.</p>

/*@<p style="font-family:verdana;font-size:15px">Either within the paragraph declaration, as is done here</p>@*/

<htest>Or in the header as is done here</htest>

<h2>Images</h2>

<p>This is how to insert an image:</p>

<img src="http://www.icosmo.org/icosmo\_logo.jpg" alt="iCosmo logo"></img>

<p>And turn an image into a hyperlink:</p>

/*@<a href="http://www.icosmo.org"><img src="http://www.icosmo.org/icosmo\_logo.jpg" alt="iCosmo logo"></img></a>@*/

<h2>Forms</h2>

/*@<p>Here is an example of a simple HTML form that will collect some information and submit it to CGI script. This CGI script will then perform a cosmological calculation using iCosmo and the information provided. (this example also shows how to make Greek letters and subscripts)</p>@*/

<p>Please enter some cosmological parameters</p>
<form action="/cgi-bin/WebPaper.cgi">
\&Omega;<sub>M</sub><input type="text" name="OmegaM" size="10" value="0.3">
\&Omega;<sub>DE</sub><input type="text" name="OmegaL" size="10" value="0.7">
<input name="reset" type="reset" value="Reset Default">
<input name="submit" type="submit" value="Calculate">
</form>

<hr>

</body>
</html>
\end{lstlisting}
}

\noindent {\bf CGI Example.} Here is the PERL code that is used to create the CGI script on {\tt http://www.icosmo.org/cgi-bin/WebPaper.cgi}, the code is also displayed at the bottom the online page. 

\noindent {\tt
\begin{lstlisting}[numbers=left]
/*@\#!/usr/bin/perl@*/
/*@require "cgi-lib.pl";@*/

/*@\&ReadParse(*input);@*/

/*@\#If not already set then set defaults:@*/
/*@\$omega\_m=0.3;@*/
/*@\$omega\_l=0.7;@*/

/*@\#Reads in the values from the webpage@*/
/*@\$omega\_m = \$input\{'OmegaM'\} if (exists \$input\{"OmegaM"\});@*/
/*@\$omega\_l = \$input\{'OmegaL'\} if (exists \$input\{"OmegaL"\});@*/

/*@print "Content-type: text/html;"$\backslash$n$\backslash$n;@*/

/*@print <<HTML;@*/
/*@<?xml version="1.0" encoding="UTF-8"?>@*/
<!DOCTYPE html PUBLIC "-//W3C//DTD XHTML 1.0 Transitional//EN" /*@\\@*/ /*@"http://www.w3.org/TR/xhtml1/DTD/xhtml1-transitional.dtd">@*/

/*@<html>@*/
/*@<head>@*/

/*@<title>A Simple Cosmology CGI Page</title>@*/

/*@</head>@*/

/*@<body>@*/

/*@<h1>Welcome to the CGI Page!</h1>@*/

/*@<p>@*/
/*@This CGI page accompanies the paper <em>Cloud Cosmology : Building the Web Interface for iCosmo</em> and has been called from the HTML page@*/ /*@\\@*/ /*@<a href="http://www.icosmo.org/WebPaper.html">http://www.icosmo.org/WebPaper.html</a>.@*/
/*@</p>@*/

/*@<p>This CGI page is written in PERL, it has read in the values you submitted. These are printed below:</p>@*/
/*@<ul>@*/
/*@<li>\&Omega;<sub>M</sub>=\$omega\_m</li>@*/
/*@<li>\&Omega;<sub>DE</sub>=\$omega\_l</li>@*/
/*@</ul>@*/

/*@<p>Now we will start an IDL session and do some very simple calculations (use the densities to determine whether the Universe is open or closed)</p>@*/
/*@HTML@*/

/*@use IPC::Open3;@*/
/*@\$ENV\{'PATH'\} = '/Applications/itt/idl70/bin/'; \$progName = "ssh wwwicosm@machine 'cd WebServer/CGI-Executables; idl'";@*/
/*@open3(WRITEIDL,READIDL,ERRORIDL, \$progName) or die "Could not begin "\$progName" ";@*/

/*@\#IDL command line arguments can now be done using the following syntax:@*/

/*@print WRITEIDL "omega\_total=\$omega\_m+\$omega\_l $\backslash$n";@*/
/*@print WRITEIDL "print,omega\_total $\backslash$n";@*/
/*@\$omega\_T=<READIDL>;@*/

/*@print "The total density is \&Omega;<sub>T</sub>=\$omega\_T so";@*/

if (\$omega\_T <= 1.0) \{
print " the Universe is open."; 
\} else \{ 
print " the Universe is closed.";
\} 

/*@print " This calculation was done in IDL. ";@*/
/*@print " Now we will use an iCosmo routine to perform a simple cosmological calculation.";@*/

/*@\#load the iCosmo routines:@*/
/*@print WRITEIDL "restore,file='comp\_fisher.sav' $\backslash$n";@*/
/*@\#set fiducial and run mk\_cosmo (without matter power spectrum):@*/
/*@print WRITEIDL "fid=set\_fiducial() $\backslash$n";@*/
/*@print WRITEIDL "fid.cosmo.omega\_m=0.3 $\backslash$n";@*/
/*@print WRITEIDL "fid.cosmo.omega\_l=0.7 $\backslash$n";@*/
/*@print WRITEIDL "cosmo=mk\_cosmo(fid,/nopk) $\backslash$n";@*/
/*@\#extract some variables:@*/
/*@print WRITEIDL "print,cosmo.evol.z[200] $\backslash$n";@*/
/*@\$z=<READIDL>;@*/
/*@print WRITEIDL "print,cosmo.evol.da[200] $\backslash$n";@*/
/*@\$da=<READIDL>;@*/

/*@print<<HTML;@*/
/*@<p style="font-family:arial;">At a redshift of \$z the angular diameter distance is \$da Mpc.</p>@*/
/*@HTML@*/

/*@\#end of the web page:@*/
/*@print<<HTML;@*/
/*@<hr>@*/
/*@</body>@*/
/*@</html>@*/
/*@HTML@*/
/*@\#close the child IDL session:@*/
/*@close(WRITEIDL);@*/
/*@close(READIDL);@*/
/*@close(ERRORIDL);@*/
/*@exit;@*/
\end{lstlisting}
}

\end{document}